\documentclass[preprint2,usegraphicx]{aastex}

\newcommand{\ms}{ms$^{-1}$}

\newcommand{\rhk}{log$R'_{\rm{HK}}$}

\shorttitle{The curious case of HD41248}
\shortauthors{Jenkins \& Tuomi}

\begin{document}



\title{The curious case of HD41248. A pair of static signals buried behind red-noise\thanks{Email: jjenkins@das.uchile.cl}}


\author{J.S. Jenkins$^{1}$ and M. Tuomi$^{1,2,3}$}
\affil{$^1$Departamento de Astronomia, Universidad de Chile, Camino el Observatorio 1515, Las Condes, Santiago, Chile, Casilla 36-D\\
$^2$Center for Astrophysics, University of Hertfordshire, College Lane Campus, Hatfield, Hertfordshire, UK, AL10 9AB\\
$^3$University of Turku, Tuorla Observatory, Department of Physics and Astronomy, V\"ais\"al\"antie 20, FI-21500, Piikki\"o, Finland}



\begin{abstract}

Gaining a better understanding of the effects of stellar induced radial velocity noise is critical for the future of exoplanet studies, since the discovery of the lowest-mass planets using this method will require us to go below the intrinsic stellar noise limit. An interesting test case in this respect is that of the southern solar analogue HD41248. The radial velocity time series of this star has been proposed to contain either a pair of signals with periods of around 18 and 25 days, that could be due to a pair of resonant super-Earths, or a single and varying 25~day signal that could arise due to a complex interplay between differential rotation and modulated activity. In this letter we build-up more evidence for the former scenario, showing that the signals are still clearly significant even after more than 10~years of observations and they likely do not change in period, amplitude, or phase as a function of time, the hallmarks of static Doppler signals. We show that over the last two observing seasons this star was more intrinsically active and the noise reddened, highlighting why better noise models are needed to find the lowest amplitude signals, in particular models that consider noise correlations.  This analysis shows that there is still sufficient evidence for the existence of two super-Earths on the edge of, or locked into, a 7:5 mean motion resonance orbiting HD41248.

\end{abstract}





\section{Introduction}

The discovery of low-mass planets in the super-Earth regime using the radial velocity method is at the forefront of modern exoplanet science as it pushes the boundaries of what is possible using current technology (\citealp{pepe11}; \citealp{tuomi14b}; \citealp{anglada-escude13}).  However, the Doppler signals imposed on the host stars of such orbiting bodies can also be fighting for dominance with signals induced in the data by rotationally modulated activity features like star spots (see \citealp{boisse11}).

A radial velocity analysis of the star HD166435 by \citet{queloz01} found a repeating short period signal of less than 4~days, suggesting the presence of a planetary companion to the star. After photometric follow-up they found a period matching the period of the radial velocity signal, indicating the star was actually active and the signal they had detected was due to rotationally modulated star spots. This was confirmed when they found that the coherence time of the radial velocity signal was only $\sim$30~days and correlations were found with the bisector inverse slope (BIS), meaning it was not a static signal as expected of a genuine Doppler velocity profile.

GJ581 provides another example of false-positive radial velocity signals where a possible candidate planet (GJ581~d) was reported in \citet{udry07} with a period of 82~days, later shown to be the 1-year alias of another planet candidate period of 67~days (\citealp{mayor09}). The existence of the habitable zone super-Earth GJ581~g (\citealp{vogt10}) has also been disputed (\citealp{tuomi12}; \citealp{baluev13}; \citealp{hatzes13b}), later countered by \citet{vogt12}, as has the existence of the Earth-mass planet reported to be orbiting Alpha Cen~B (\citealp{dumusque12}; \citealp{hatzes13}). Clearly the detection of low-mass planets approaching the intrinsic noise level of the star and instrument combination is fraught with difficulty.

\citet{jenkins13b} announced the discovery of a pair of planetary candidates orbiting the star HD41248 in, or close to a 7:5 mean motion resonance (MMR) configuration. Both signals reported in their work were statistically significant, even when considering correlations between the activity indicators and the radial velocities. However, although the time baseline was long, around 7.5~years, they only had a total of 62 Doppler velocities, yet the MMR configuration (period ratio of 1.400$\pm$0.002) seemed to favour a planetary hypothesis as such a pair of periods so close to a 7:5 integer ratio seems difficult to attribute to the star. The metal-poor nature of HD41248 ([Fe/H]=-0.43~dex) also agrees with the emerging notion that metal-poor stars have a higher fraction of the lowest-mass planets (\citealp{jenkins13a}). 

Recently, \citet{santos14} have claimed that the longer period signal in the HD41248 radial velocity data is likely due to rotationally modulated magnetic activity, after adding more than 160 new velocities, and when subtracting off that signal, there is no remaining evidence for the 18~day signal. With this in mind we decided to re-analyze all data for HD41248 and test whether the pair of signals still remain in the new data from \citeauthor{santos14}. Moreover, we discuss whether these signals could still be interpreted as being due to a pair of planets.

\section{HD41248 Statistical Model}

We modelled the HARPS radial velocities of HD42148 by adopting the analysis techniques and the statistical model applied in \citet{tuomi14c}. This model contains Keplerian signals, a linear trend, moving average component with exponential smoothing, and linear correlations with activity indices, namely, BIS, full width at half maximum (FWHM), and chromospheric activity $S$-index. According to \citeauthor{tuomi14c} such a model can filter out activity-related variations in radial velocities and even suppress the velocity variations caused by the co-rotation of star spots on the stellar surface below the detection threshold, enabling the detection of low-amplitude variations of planetary origin, as witnessed on CoRoT-7 (\citeauthor{tuomi14c}).  We write the statistical model as

\begin{eqnarray}\label{eq:model}
  m_{i,l} = \gamma_{l} + \dot{\gamma}t_{i} + f_{k}(t_{i}) + \epsilon_{i,l}  + \sum_{j=1}^{q} c_{j,l} \xi_{j,i,l} \nonumber\\
  + \sum_{j=1}^{p} \phi_{j,l} \exp \Bigg\{ \frac{t_{i-j} - t_{i}}{\tau_{l}} \Bigg\} \epsilon_{i,l} ,
\end{eqnarray}

where $m_{i,l}$ is the measurement made at time $t_{i}$ and the index $l$ denotes that it corresponds to an independent $l$th data set, parameter $\gamma_{l}$ is the reference velocity, function $f_{k}$ denotes the superposition of $k$ Keplerian signals, $\epsilon_{i,l}$ is a Gaussian white noise with zero mean and a variance of $\sigma_{i}^{2} + \sigma_{l}^{2}$, where $\sigma_{i}$ is the estimated instrument uncertainty corresponding to the radial velocity measurement $m_{i,l}$ and $\sigma_{l}$ quantifies the excess white noise in the $l$th data set, parameters $c_{j,l}$ describe the linear correlations with the activity indices $\xi_{j,i,l}$, for $j=1, ..., q$, and parameters $\phi_{j,l}$ quantify the moving average components, $j=1, ..., p$, with exponential smoothing in a time-scale of $\tau_{l}$.  In practice we apply a first-order moving average model (MA(1)) as we believe it is a sufficiently accurate description of this data, parameterised by setting the moving average components ($p$) in Eq$^{n}$~\ref{eq:model} equal to unity.  

The prior probability densities have to be defined in order to use the techniques of \citet{tuomi14c} relying on the Bayes' rule of conditional probabilities. We define these densities according to \citet{tuomi13c} by choosing uninformative and uniform densities for all but two model parameters, namely the eccentricities ($e$) and excess jitter ($\sigma_{\rm J}$). These are set such that low eccentricities and low jitters are preferred but that higher values are not ruled out \emph{a priori} \citep{tuomi14c}.

As the noise caused by inhomogeneities of the stellar surface and activity cannot be expected to be time-invariant over the baseline of the observations of over ten years, we model the velocities already analysed in \citep{jenkins13b} and the new ones obtained during the last two years as independent data sets.  In this way, we can account for the possibility that the noise properties have changed over the data baseline and the potential effects that the data sampling, which is more dense during the last two years, has on the parameters of the noise model.  Finally, we also split the last two observing seasons up into two subsets of data, and although this is detrimental to the information content, this was done to directly compare our results with those recently published in \citet{santos14}. 

\section{HD41248 Reanalysis}

We applied our statistical model outlined above to the full dataset of radial velocities for HD41248, combining the previously published data in \citet{jenkins13b} with the newly published data in \citet{santos14}, giving rise to a total timeseries of 223 HARPS (\citealp{mayor03b}) velocities\footnote{The data were obtained from the European Southern Observatory archive under the request number JJENKINS-110394.}. We applied both tests with and without including the linear activity correlation terms and also compared to the white noise model applied in \citeauthor{santos14}  Table~1 contains the measured radial velocities, BIS and FWHM values from the HARPS CCFs, and the chromospheric $S$-indices that were measured following the procedures in \citet{jenkins06,jenkins08,jenkins11}.

\begin{deluxetable}{ccccc}
\center
\tablecaption{HARPS timeseries data for HD41248.}
\label{rv_data}
\tablehead{
\colhead{BJD} & \colhead{RV [\ms]} & \colhead{$S$-index [dex]} & \colhead{BIS [\ms]} & \colhead{FWHM [\ms]} }
\startdata
2452943.8528426 &  3526.59$\pm$2.59 & 0.169 & 35.93 & 6721.78 \\
2452989.7102293 &  3519.14$\pm$4.06 & 0.170 & 27.40 & 6719.01\\
2452998.6898180 &  3526.43$\pm$5.43 & 0.179 & 33.53 & 6701.21\\
2453007.6786518 &  3526.63$\pm$2.53 & 0.162 & 28.61 & 6718.20\\
2453787.6079555 &  3522.44$\pm$2.76 & 0.162 & 31.31 & 6718.54\\
2454055.8375443 &  3523.18$\pm$2.06 & 0.168 & 23.95 & 6714.52\\
2454789.7207967 &  3522.99$\pm$0.82 & 0.171 & 27.43 & 6722.19\\
2454790.6943362 &  3519.49$\pm$0.90 & 0.170 & 30.83 & 6724.20\\
2454791.7055725 &  3522.47$\pm$0.83 & 0.171 & 29.54 & 6720.60\\
2454792.7042506 &  3522.29$\pm$0.80 & 0.172 & 28.09 & 6728.65\\
2454793.7211230 &  3524.99$\pm$0.89 & 0.173 & 25.28 & 6727.73\\
2454794.6946036 &  3527.04$\pm$0.89 & 0.172 & 29.59 & 6732.04\\
2454795.7156306 &  3528.45$\pm$0.91 & 0.174 & 29.54 & 6725.92\\
2454796.7195391 &  3528.21$\pm$0.96 & 0.174 & 30.33 & 6727.87\\
2454797.7051254 &  3528.99$\pm$0.91 & 0.175 & 27.77 & 6733.22\\
2454798.6972277 &  3531.20$\pm$0.92 & 0.173 & 25.14 & 6731.22\\
2454902.5907553 &  3525.08$\pm$2.02 & 0.180 & 21.34 & 6729.23\\
2454903.5172666 &  3527.59$\pm$0.78 & 0.172 & 27.69 & 6726.66\\
2454904.5185682 &  3525.76$\pm$0.90 & 0.174 & 27.19 & 6722.67\\
2454905.5355291 &  3527.55$\pm$0.90 & 0.171 & 27.98 & 6722.94\\
2454906.5179999 &  3527.95$\pm$1.04 & 0.171 & 31.02 & 6723.86\\
2454907.5647983 &  3527.52$\pm$0.91 & 0.173 & 28.88 & 6727.05\\
2454908.5603822 &  3526.23$\pm$0.80 & 0.172 & 27.95 & 6720.84\\
2454909.5380036 &  3527.11$\pm$0.85 & 0.170 & 25.45 & 6721.36\\
2454910.5385064 &  3528.24$\pm$1.12 & 0.173 & 27.51 & 6726.82\\
2454911.5427244 &  3524.98$\pm$0.75 & 0.172 & 28.13 & 6721.05\\
2454912.5392921 &  3525.22$\pm$0.74 & 0.172 & 25.78 & 6718.78\\
2455284.5272133 &  3528.23$\pm$0.73 & 0.175 & 28.43 & 6724.46\\
2455287.5109103 &  3524.60$\pm$0.88 & 0.174 & 25.19 & 6733.01\\
2455288.5285775 &  3523.32$\pm$0.75 & 0.173 & 25.44 & 6730.41\\
2455289.5460248 &  3526.70$\pm$0.79 & 0.174 & 26.29 & 6723.60\\
2455290.5095380 &  3525.90$\pm$0.89 & 0.174 & 21.96 & 6727.32\\
2455291.5216615 &  3526.08$\pm$1.01 & 0.171 & 26.86 & 6734.67\\
2455293.5043818 &  3527.36$\pm$0.97 & 0.172 & 27.39 & 6727.22\\
2455304.5180173 &  3522.41$\pm$1.69 & 0.144 & 31.21 & 6801.23\\
2455328.4550220 &  3529.11$\pm$0.79 & 0.175 & 20.74 & 6735.39\\
2455334.4564390 &  3532.00$\pm$1.37 & 0.169 & 25.33 & 6737.32\\
2455387.9305071 &  3530.98$\pm$1.04 & 0.171 & 27.29 & 6737.65\\
2455390.9312188 &  3531.33$\pm$1.58 & 0.164 & 30.24 & 6734.98\\
2455434.8790630 &  3516.75$\pm$3.33 & 0.157 & 31.59 & 6740.53\\
2455439.8843407 &  3527.76$\pm$3.26 & 0.159 & 32.81 & 6723.48\\
2455445.9241644 &  3522.93$\pm$3.11 & 0.162 & 28.31 & 6727.79\\
2455465.8566225 &  3526.62$\pm$1.41 & 0.171 & 23.18 & 6731.85\\
2455480.8795598 &  3527.59$\pm$1.13 & 0.171 & 27.10 & 6728.47\\
2455483.8136024 &  3525.01$\pm$1.74 & 0.170 & 27.97 & 6734.02\\
2455488.8262312 &  3525.84$\pm$0.75 & 0.171 & 31.08 & 6725.48\\
2455494.8532009 &  3529.81$\pm$0.95 & 0.169 & 26.88 & 6730.95\\
2455513.7823100 &  3528.86$\pm$1.29 & 0.172 & 35.60 & 6740.83\\
2455516.7515789 &  3530.74$\pm$0.94 & 0.175 & 22.72 & 6736.79\\
2455519.7046533 &  3528.85$\pm$1.20 & 0.168 & 21.96 & 6735.81\\
2455537.7997291 &  3527.71$\pm$0.72 & 0.176 & 27.01 & 6730.30\\
2455545.7213656 &  3529.20$\pm$0.85 & 0.175 & 26.69 & 6737.60\\
2455549.7548559 &  3527.95$\pm$0.83 & 0.171 & 31.98 & 6734.55\\
2455576.7923481 &  3525.35$\pm$1.01 & 0.165 & 29.18 & 6733.68\\
2455580.7312518 &  3519.41$\pm$0.90 & 0.169 & 32.95 & 6729.42\\
2455589.7734088 &  3528.42$\pm$1.42 & 0.169 & 24.59 & 6733.73\\
2455612.6068850 &  3527.66$\pm$0.82 & 0.171 & 25.59 & 6731.09\\
2455623.6361828 &  3528.30$\pm$1.18 & 0.169 & 28.96 & 6726.23\\
2455629.5528393 &  3528.65$\pm$0.94 & 0.168 & 25.71 & 6732.35\\
2455641.5542106 &  3528.98$\pm$1.01 & 0.170 & 28.91 & 6727.21\\
2455644.5845033 &  3529.35$\pm$1.26 & 0.174 & 19.09 & 6740.11\\
2455647.5796694 &  3529.95$\pm$1.37 & 0.167 & 29.11 & 6737.56\\
2455904.8445150 &  3523.18$\pm$2.58 & 0.158 & 40.69 & 6747.45\\
2456215.8328718 &  3526.25$\pm$1.39 & 0.168 & 30.20 & 6743.55\\
2456218.7962306 &  3524.50$\pm$1.84 & 0.164 & 28.94 & 6740.25\\
2456218.8693916 &  3523.75$\pm$1.62 & 0.165 & 35.11 & 6739.30\\
2456220.8683473 &  3524.85$\pm$1.09 & 0.163 & 25.77 & 6731.64\\
2456229.7183898 &  3532.26$\pm$1.69 & 0.166 & 31.33 & 6740.83\\
2456229.8714013 &  3527.06$\pm$1.40 & 0.166 & 29.16 & 6736.86\\
2456230.6857996 &  3530.13$\pm$1.47 & 0.167 & 27.20 & 6743.88\\
2456230.8458134 &  3531.97$\pm$1.34 & 0.167 & 28.59 & 6736.59\\
2456231.7334650 &  3531.02$\pm$1.55 & 0.171 & 29.46 & 6749.04\\
2456231.8607353 &  3529.80$\pm$1.20 & 0.168 & 28.35 & 6741.43\\
2456236.6604420 &  3524.84$\pm$1.51 & 0.173 & 29.56 & 6745.23\\
2456236.8633497 &  3524.75$\pm$1.19 & 0.174 & 33.24 & 6733.75\\
2456237.6662421 &  3525.98$\pm$1.33 & 0.177 & 28.27 & 6738.88\\
2456237.8808398 &  3525.49$\pm$0.90 & 0.176 & 26.42 & 6741.69\\
2456238.6619956 &  3532.88$\pm$1.31 & 0.175 & 29.30 & 6746.51\\
2456238.6977368 &  3527.82$\pm$1.33 & 0.173 & 28.61 & 6742.04\\
2456239.6522975 &  3532.63$\pm$1.42 & 0.175 & 24.85 & 6747.45\\
2456245.7084222 &  3525.35$\pm$1.09 & 0.173 & 24.69 & 6743.09\\
2456247.6418575 &  3528.74$\pm$1.02 & 0.168 & 28.32 & 6741.03\\
2456256.6533194 &  3532.39$\pm$2.19 & 0.167 & 24.63 & 6747.43\\
2456256.7645704 &  3525.52$\pm$1.24 & 0.175 & 25.11 & 6732.43\\
2456257.6680498 &  3533.01$\pm$1.16 & 0.173 & 23.79 & 6744.10\\
2456257.8001458 &  3530.94$\pm$1.23 & 0.178 & 32.49 & 6743.71\\
2456258.7216343 &  3527.87$\pm$1.51 & 0.171 & 19.43 & 6747.10\\
2456258.7556855 &  3529.95$\pm$1.47 & 0.173 & 29.98 & 6746.68\\
2456259.7335053 &  3524.76$\pm$1.30 & 0.175 & 23.97 & 6746.34\\
2456259.7520008 &  3525.54$\pm$1.26 & 0.175 & 23.55 & 6745.35\\
2456262.6948682 &  3529.96$\pm$1.62 & 0.176 & 25.69 & 6742.39\\
2456262.7981443 &  3523.17$\pm$1.32 & 0.178 & 27.46 & 6732.01\\
2456263.6232652 &  3532.22$\pm$1.51 & 0.174 & 26.24 & 6741.81\\
2456263.8103729 &  3529.64$\pm$1.14 & 0.174 & 24.40 & 6748.01\\
2456307.5801789 &  3532.77$\pm$1.00 & 0.171 & 27.54 & 6743.34\\
2456307.7092172 &  3533.96$\pm$1.33 & 0.180 & 24.02 & 6749.01\\
2456308.5535163 &  3530.95$\pm$1.08 & 0.176 & 28.15 & 6736.73\\
2456308.8012459 &  3533.41$\pm$1.41 & 0.170 & 24.73 & 6739.00\\
2456309.5762513 &  3533.01$\pm$1.10 & 0.178 & 28.63 & 6744.66\\
2456309.7930203 &  3531.16$\pm$1.28 & 0.172 & 31.86 & 6747.49\\
2456310.5910230 &  3532.02$\pm$1.19 & 0.177 & 18.72 & 6741.44\\
2456310.7069248 &  3536.34$\pm$1.26 & 0.177 & 19.30 & 6738.18\\
2456311.5625541 &  3532.26$\pm$1.34 & 0.171 & 28.39 & 6747.14\\
2456311.6578889 &  3530.72$\pm$1.20 & 0.176 & 28.15 & 6742.01\\
2456312.5914918 &  3529.65$\pm$1.26 & 0.174 & 26.59 & 6741.52\\
2456312.7438284 &  3534.54$\pm$2.32 & 0.169 & 17.33 & 6759.18\\
2456314.5898762 &  3531.61$\pm$1.61 & 0.176 & 24.95 & 6757.65\\
2456314.7537057 &  3530.93$\pm$1.62 & 0.173 & 28.79 & 6755.44\\
2456315.6175515 &  3528.47$\pm$2.03 & 0.191 & 31.34 & 6756.00\\
2456315.7297379 &  3528.90$\pm$1.58 & 0.184 & 23.65 & 6749.19\\
2456316.6766846 &  3532.46$\pm$1.57 & 0.182 & 19.56 & 6745.99\\
2456317.5901359 &  3526.93$\pm$1.14 & 0.173 & 23.48 & 6745.87\\
2456317.6744291 &  3530.39$\pm$1.27 & 0.171 & 21.48 & 6743.95\\
2456318.5677531 &  3529.95$\pm$1.37 & 0.173 & 28.49 & 6741.46\\
2456318.6886662 &  3535.64$\pm$2.18 & 0.185 & 26.26 & 6752.97\\
2456319.5818048 &  3526.20$\pm$1.52 & 0.178 & 26.31 & 6736.50\\
2456319.7199515 &  3517.28$\pm$2.59 & 0.178 & 36.77 & 6759.41\\
2456320.6218166 &  3526.41$\pm$1.13 & 0.174 & 32.82 & 6739.65\\
2456320.7117346 &  3526.74$\pm$1.60 & 0.172 & 23.78 & 6736.66\\
2456321.6454163 &  3527.10$\pm$1.74 & 0.174 & 26.13 & 6744.54\\
2456321.7454383 &  3526.19$\pm$1.56 & 0.174 & 27.20 & 6744.40\\
2456326.5925681 &  3526.19$\pm$1.65 & 0.172 & 32.18 & 6744.18\\
2456326.6716412 &  3524.07$\pm$1.65 & 0.178 & 25.37 & 6736.35\\
2456354.5410577 &  3527.23$\pm$1.13 & 0.165 & 32.33 & 6731.94\\
2456354.6178620 &  3522.61$\pm$1.82 & 0.169 & 26.08 & 6735.38\\
2456357.5276034 &  3529.82$\pm$1.59 & 0.166 & 29.62 & 6745.95\\
2456357.6455296 &  3530.10$\pm$1.42 & 0.163 & 23.82 & 6742.41\\
2456362.5180018 &  3528.99$\pm$1.14 & 0.175 & 27.93 & 6737.80\\
2456362.6121205 &  3528.42$\pm$1.28 & 0.177 & 25.45 & 6744.41\\
2456366.5017386 &  3526.21$\pm$0.98 & 0.175 & 24.63 & 6731.86\\
2456366.6539234 &  3527.18$\pm$2.20 & 0.174 & 25.56 & 6758.16\\
2456383.5228456 &  3527.32$\pm$1.81 & 0.173 & 27.84 & 6743.47\\
2456384.5783826 &  3531.49$\pm$1.66 & 0.163 & 17.53 & 6734.83\\
2456385.4956349 &  3526.67$\pm$1.34 & 0.165 & 25.32 & 6732.94\\
2456385.5663744 &  3529.43$\pm$1.84 & 0.168 & 30.42 & 6741.01\\
2456386.4894948 &  3529.38$\pm$1.39 & 0.168 & 24.65 & 6741.94\\
2456386.5607204 &  3527.55$\pm$1.25 & 0.164 & 22.69 & 6737.07\\
2456387.5267452 &  3527.35$\pm$1.27 & 0.180 & 26.54 & 6738.97\\
2456387.5630872 &  3526.19$\pm$1.49 & 0.170 & 28.71 & 6742.92\\
2456389.4913755 &  3530.50$\pm$1.22 & 0.170 & 28.87 & 6737.10\\
2456389.5578790 &  3529.82$\pm$1.34 & 0.168 & 30.95 & 6741.93\\
2456390.4892403 &  3528.62$\pm$1.35 & 0.170 & 27.67 & 6740.02\\
2456390.5645400 &  3527.94$\pm$1.32 & 0.170 & 25.35 & 6743.67\\
2456394.5131767 &  3520.93$\pm$1.34 & 0.166 & 29.75 & 6742.50\\
2456397.5056966 &  3522.49$\pm$1.36 & 0.165 & 28.01 & 6737.56\\
2456399.5231059 &  3524.07$\pm$1.05 & 0.172 & 30.15 & 6735.54\\
2456402.5030123 &  3528.33$\pm$1.06 & 0.168 & 30.25 & 6739.08\\
2456409.4955108 &  3530.07$\pm$7.47 & 0.205 & 39.30 & 6717.70\\
2456414.4684733 &  3532.52$\pm$1.19 & 0.170 & 23.77 & 6743.40\\
2456521.9016581 &  3525.91$\pm$1.32 & 0.168 & 26.89 & 6749.61\\
2456524.9204178 &  3527.03$\pm$0.92 & 0.174 & 31.13 & 6744.43\\
2456525.9077938 &  3533.12$\pm$2.05 & 0.165 & 22.16 & 6744.49\\
2456526.9330524 &  3525.49$\pm$1.04 & 0.170 & 30.36 & 6745.33\\
2456534.9211421 &  3525.52$\pm$0.85 & 0.169 & 26.82 & 6736.37\\
2456538.8521797 &  3536.99$\pm$1.65 & 0.167 & 29.80 & 6750.23\\
2456538.9215906 &  3530.71$\pm$1.35 & 0.171 & 25.16 & 6741.72\\
2456539.9012946 &  3532.42$\pm$1.59 & 0.174 & 28.36 & 6747.01\\
2456542.9158485 &  3529.81$\pm$1.42 & 0.170 & 33.66 & 6748.92\\
2456543.9208079 &  3531.25$\pm$1.08 & 0.174 & 24.88 & 6748.21\\
2456564.7508993 &  3531.65$\pm$1.78 & 0.166 & 26.35 & 6745.64\\
2456564.9043048 &  3534.39$\pm$1.82 & 0.166 & 25.79 & 6751.38\\
2456565.7808829 &  3532.73$\pm$0.93 & 0.171 & 30.83 & 6747.99\\
2456565.8439396 &  3531.39$\pm$1.82 & 0.168 & 27.50 & 6746.86\\
2456585.7655987 &  3524.51$\pm$1.20 & 0.171 & 27.67 & 6746.72\\
2456585.8490838 &  3523.68$\pm$1.04 & 0.171 & 33.58 & 6746.03\\
2456586.7691910 &  3530.26$\pm$1.61 & 0.169 & 25.38 & 6746.28\\
2456586.8643661 &  3528.84$\pm$1.11 & 0.173 & 29.25 & 6738.47\\
2456589.8651653 &  3531.07$\pm$0.92 & 0.175 & 33.10 & 6740.17\\
2456590.7783506 &  3529.10$\pm$1.47 & 0.170 & 31.50 & 6751.98\\
2456590.8722988 &  3533.43$\pm$1.38 & 0.168 & 25.97 & 6756.39\\
2456591.7566178 &  3532.10$\pm$1.58 & 0.175 & 23.82 & 6751.30\\
2456591.8177645 &  3531.08$\pm$1.27 & 0.174 & 26.57 & 6754.02\\
2456592.7146993 &  3534.20$\pm$1.23 & 0.174 & 26.47 & 6751.38\\
2456592.8521782 &  3533.31$\pm$1.03 & 0.178 & 24.86 & 6743.74\\
2456593.8707581 &  3532.37$\pm$1.36 & 0.174 & 32.43 & 6749.38\\
2456594.7546831 &  3531.44$\pm$1.46 & 0.169 & 32.42 & 6753.41\\
2456594.8060264 &  3527.69$\pm$1.57 & 0.174 & 34.27 & 6748.10\\
2456596.6972805 &  3527.60$\pm$1.32 & 0.173 & 24.39 & 6762.69\\
2456596.8585212 &  3526.84$\pm$1.75 & 0.170 & 27.56 & 6747.78\\
2456599.7447303 &  3526.40$\pm$1.04 & 0.171 & 30.20 & 6743.89\\
2456599.8591416 &  3524.32$\pm$1.34 & 0.171 & 23.41 & 6745.79\\
2456600.8594907 &  3525.48$\pm$1.03 & 0.170 & 25.14 & 6741.79\\
2456601.7290393 &  3528.45$\pm$0.92 & 0.173 & 28.60 & 6746.62\\
2456601.8378834 &  3525.90$\pm$0.79 & 0.169 & 27.63 & 6739.36\\
2456602.7714141 &  3525.90$\pm$1.04 & 0.168 & 32.24 & 6742.33\\
2456604.8350746 &  3525.25$\pm$0.96 & 0.169 & 31.53 & 6739.14\\
2456608.8439887 &  3523.29$\pm$1.30 & 0.172 & 27.65 & 6743.17\\
2456610.6894263 &  3522.93$\pm$1.07 & 0.169 & 27.50 & 6739.51\\
2456610.8735256 &  3521.37$\pm$1.10 & 0.171 & 30.97 & 6737.86\\
2456612.7502127 &  3527.77$\pm$1.10 & 0.174 & 33.52 & 6740.98\\
2456612.8583272 &  3526.78$\pm$1.12 & 0.172 & 29.80 & 6738.38\\
2456613.6797479 &  3529.07$\pm$1.22 & 0.169 & 28.03 & 6735.98\\
2456613.8119367 &  3526.60$\pm$1.27 & 0.171 & 24.17 & 6743.79\\
2456614.6882999 &  3531.03$\pm$0.93 & 0.170 & 27.78 & 6748.94\\
2456614.8534636 &  3530.97$\pm$0.88 & 0.170 & 26.26 & 6742.29\\
2456616.7207968 &  3536.57$\pm$1.04 & 0.175 & 31.38 & 6749.62\\
2456616.8313997 &  3534.77$\pm$1.01 & 0.174 & 28.86 & 6745.30\\
2456617.6567663 &  3535.31$\pm$1.15 & 0.176 & 29.28 & 6748.42\\
2456617.7938856 &  3535.82$\pm$1.13 & 0.175 & 27.73 & 6750.32\\
2456618.6626320 &  3540.55$\pm$1.05 & 0.178 & 29.50 & 6751.19\\
2456618.7062435 &  3538.81$\pm$0.94 & 0.173 & 28.45 & 6754.25\\
2456619.7046552 &  3532.96$\pm$1.00 & 0.174 & 27.55 & 6745.56\\
2456619.7752114 &  3530.89$\pm$1.03 & 0.177 & 29.71 & 6753.46\\
2456620.6982402 &  3528.84$\pm$1.51 & 0.169 & 27.96 & 6751.00\\
2456620.7760071 &  3531.28$\pm$1.37 & 0.174 & 33.57 & 6756.66\\
2456622.6838006 &  3526.40$\pm$1.42 & 0.169 & 19.42 & 6756.13\\
2456623.7458227 &  3525.46$\pm$0.98 & 0.174 & 24.04 & 6742.54\\
2456623.8656385 &  3525.04$\pm$1.03 & 0.170 & 23.45 & 6748.55\\
2456624.6414198 &  3524.73$\pm$1.27 & 0.171 & 29.66 & 6741.04\\
2456624.7668605 &  3522.75$\pm$1.17 & 0.169 & 34.98 & 6733.78\\
2456626.6156583 &  3525.44$\pm$1.25 & 0.175 & 29.43 & 6741.58\\
2456626.8487502 &  3523.88$\pm$1.10 & 0.170 & 28.93 & 6739.69\\
2456627.6233271 &  3524.30$\pm$1.10 & 0.168 & 30.72 & 6742.43\\
2456627.8381896 &  3524.37$\pm$0.89 & 0.169 & 24.09 & 6740.67\\
2456628.6101737 &  3524.76$\pm$2.28 & 0.166 & 27.97 & 6745.50\\
2456628.8421543 &  3526.22$\pm$1.49 & 0.174 & 25.51 & 6739.56\\
2456629.6267424 &  3526.08$\pm$1.74 & 0.165 & 34.21 & 6744.21\\
2456629.8238847 &  3526.80$\pm$1.21 & 0.171 & 32.54 & 6740.69\\
2456630.6085999 &  3527.11$\pm$1.47 & 0.168 & 36.99 & 6746.40\\
2456631.6333972 &  3527.87$\pm$1.02 & 0.167 & 31.01 & 6747.34\\
2456631.8450650 &  3523.59$\pm$1.08 & 0.168 & 29.85 & 6739.82\\
2456632.6133791 &  3526.43$\pm$0.98 & 0.172 & 26.77 & 6744.59\\
2456632.8576512 &  3523.42$\pm$1.14 & 0.168 & 23.92 & 6744.24\\
\enddata
\end{deluxetable}

\begin{figure*}

\includegraphics[angle=270, width=0.32\textwidth, clip]{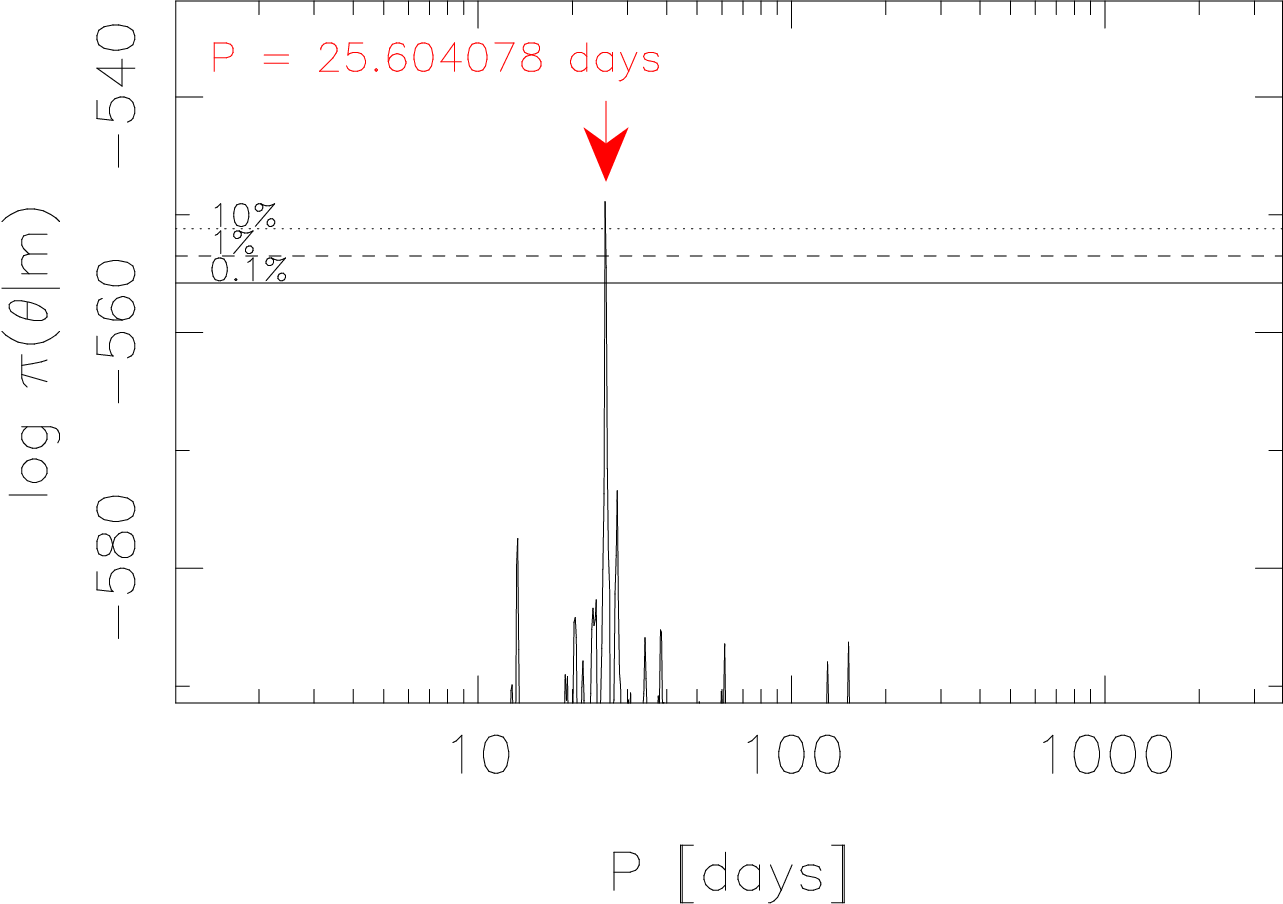}
\includegraphics[angle=270, width=0.32\textwidth, clip]{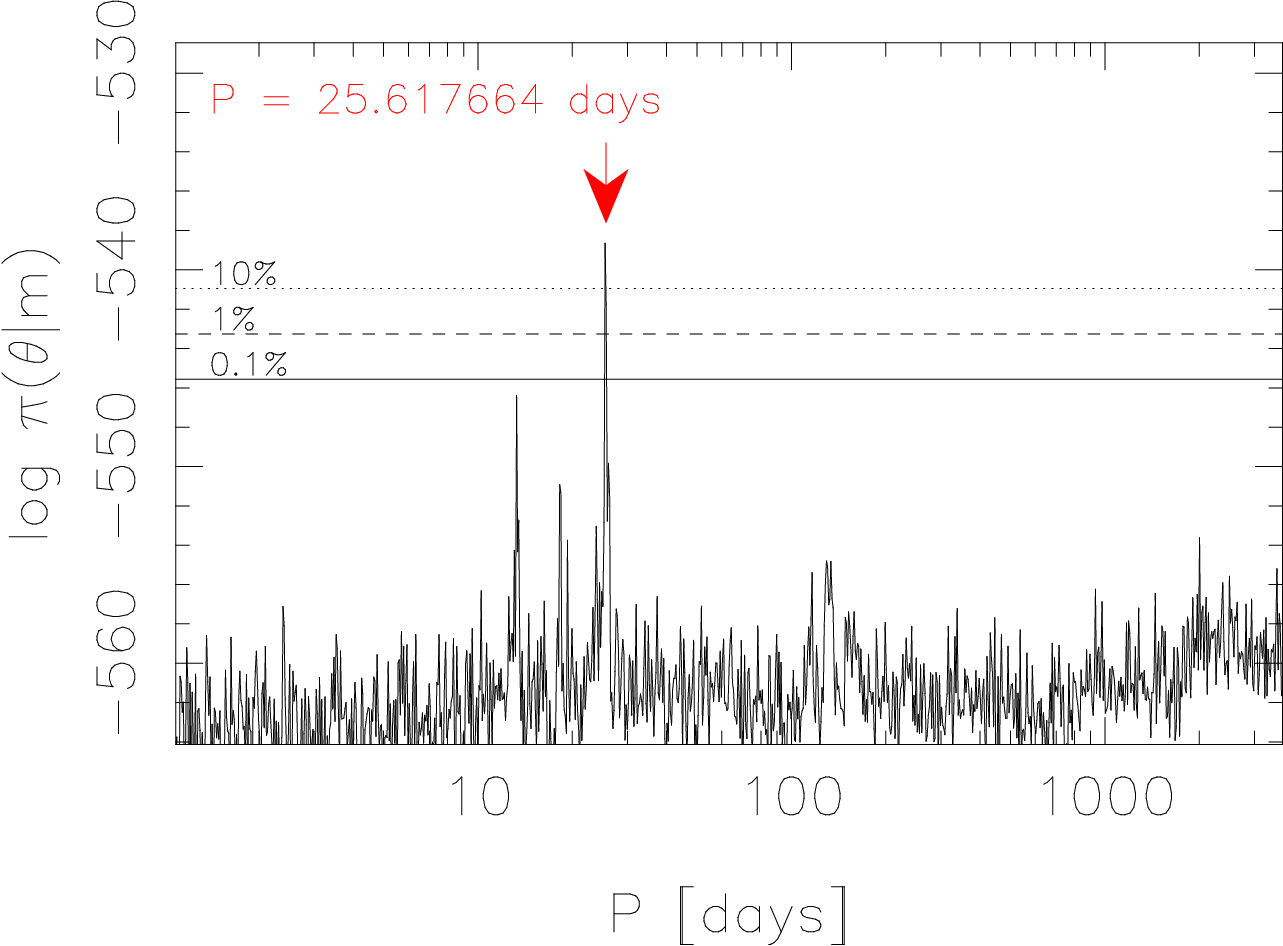}
\includegraphics[angle=270, width=0.32\textwidth, clip]{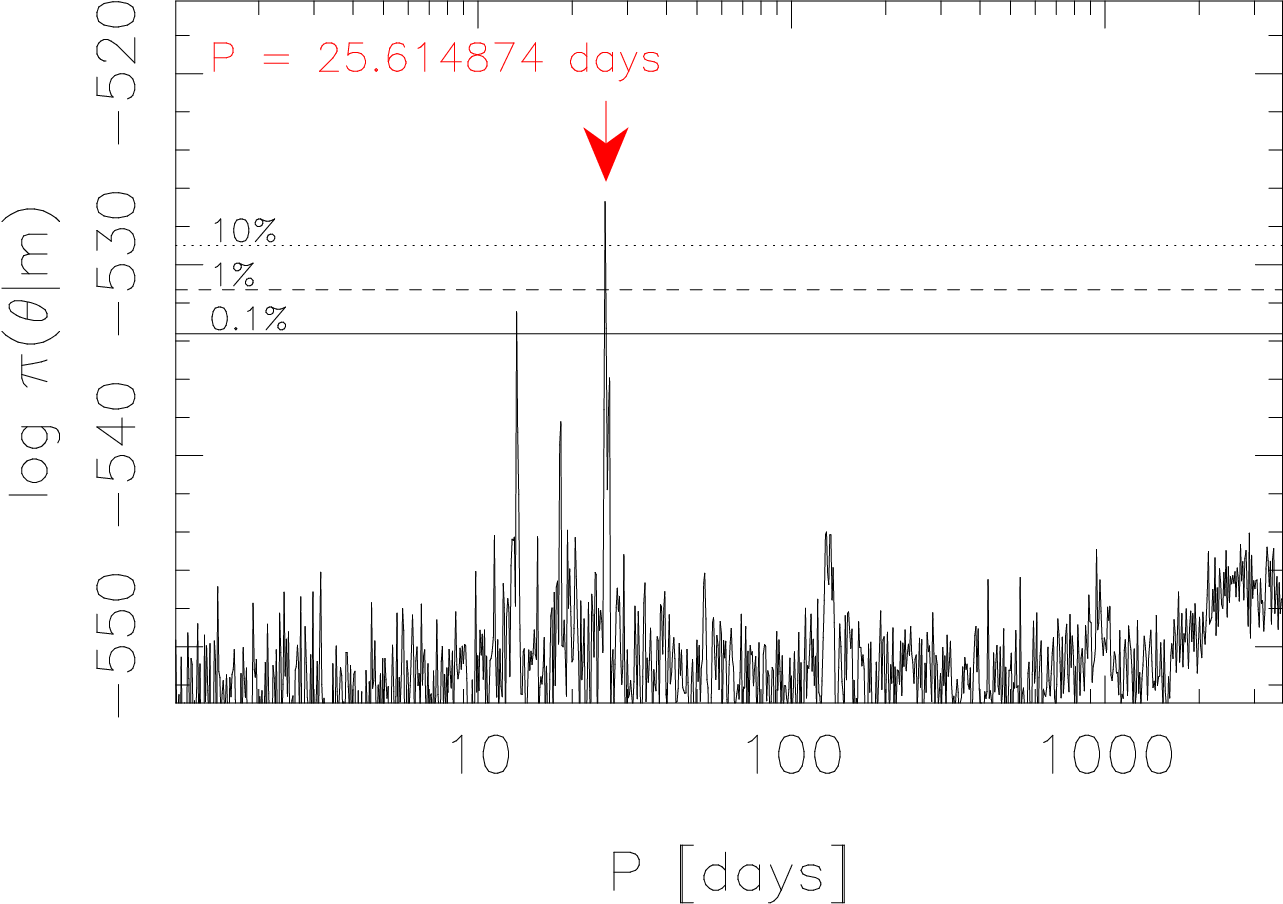}

\includegraphics[angle=270, width=0.32\textwidth, clip]{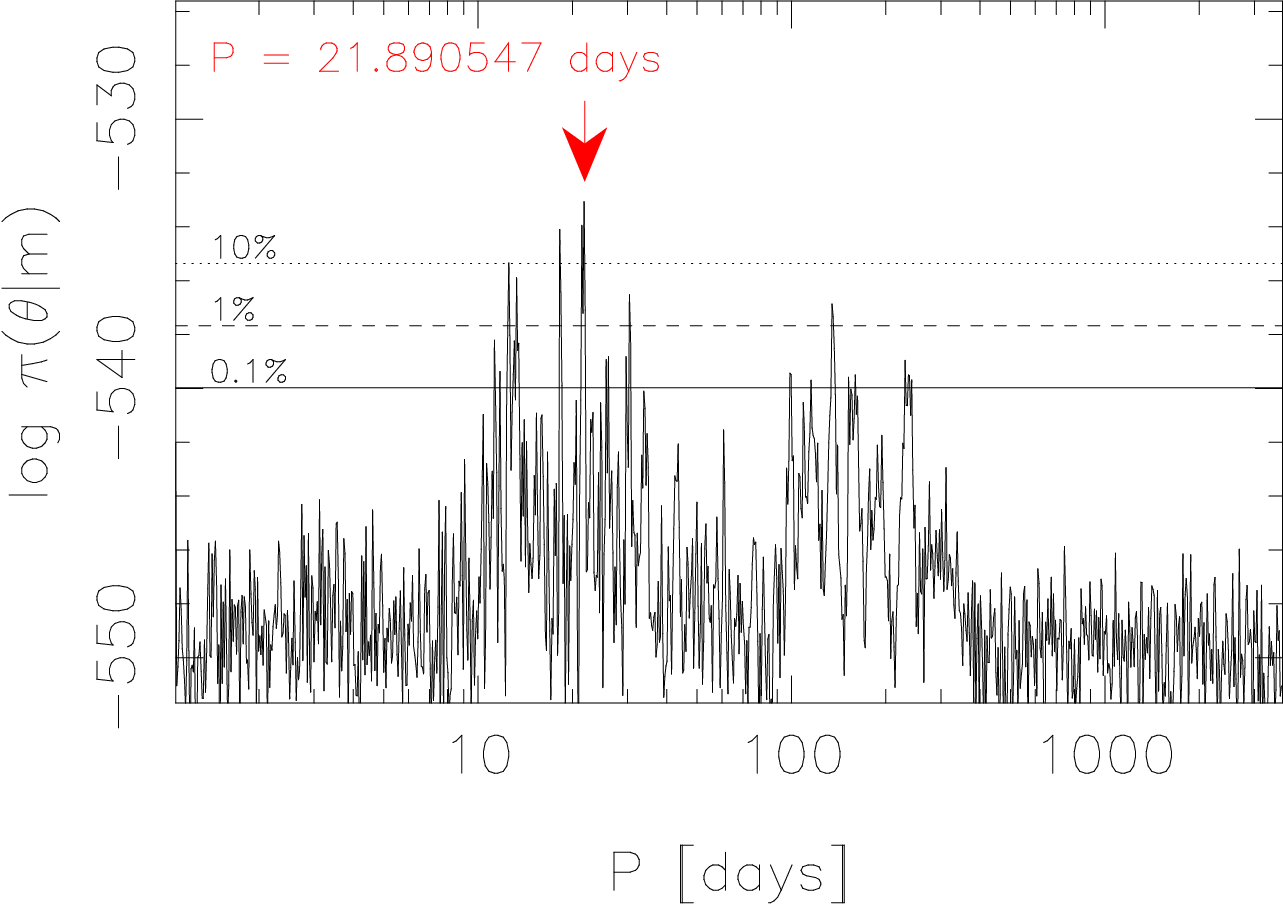}
\includegraphics[angle=270, width=0.32\textwidth, clip]{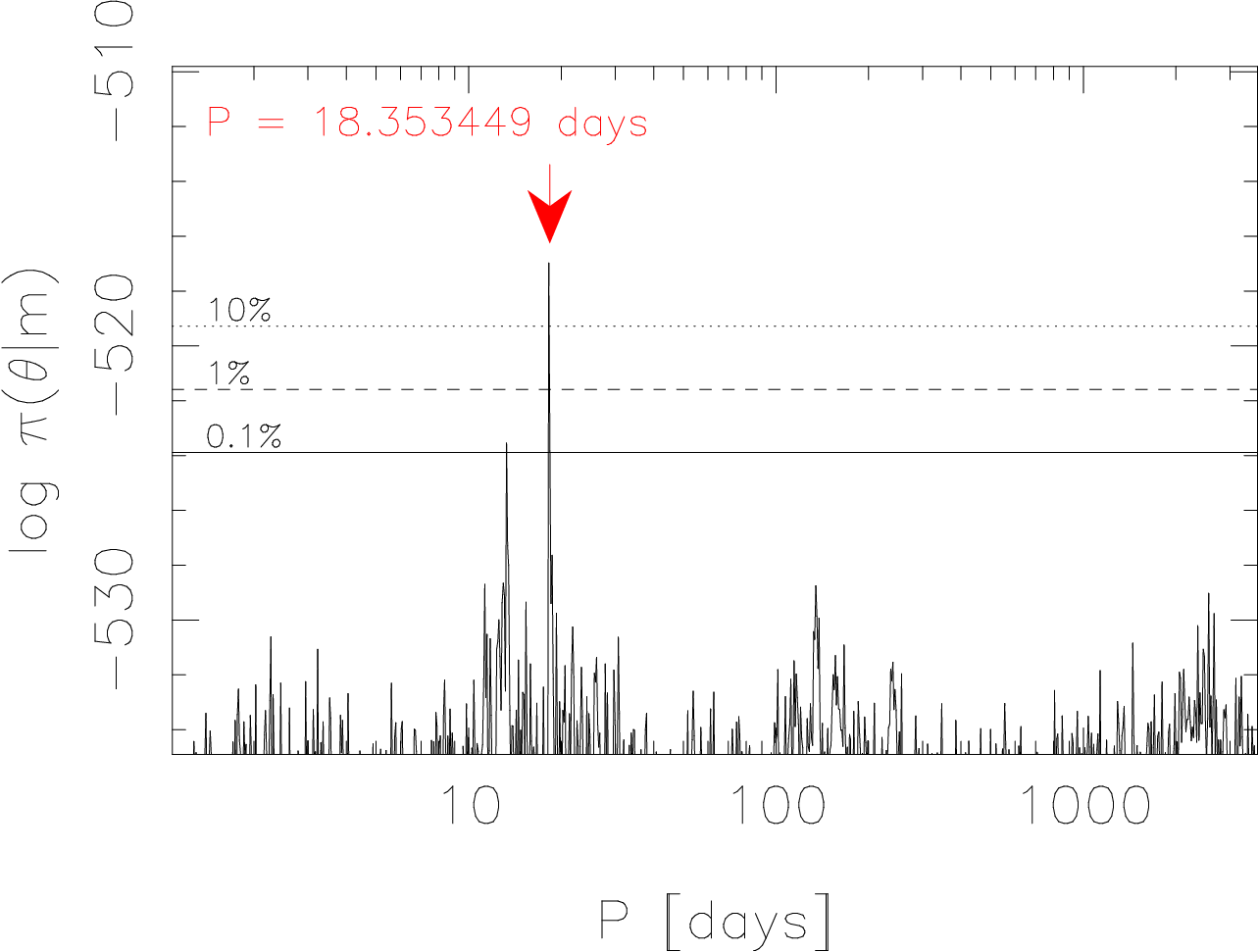}
\includegraphics[angle=270, width=0.32\textwidth, clip]{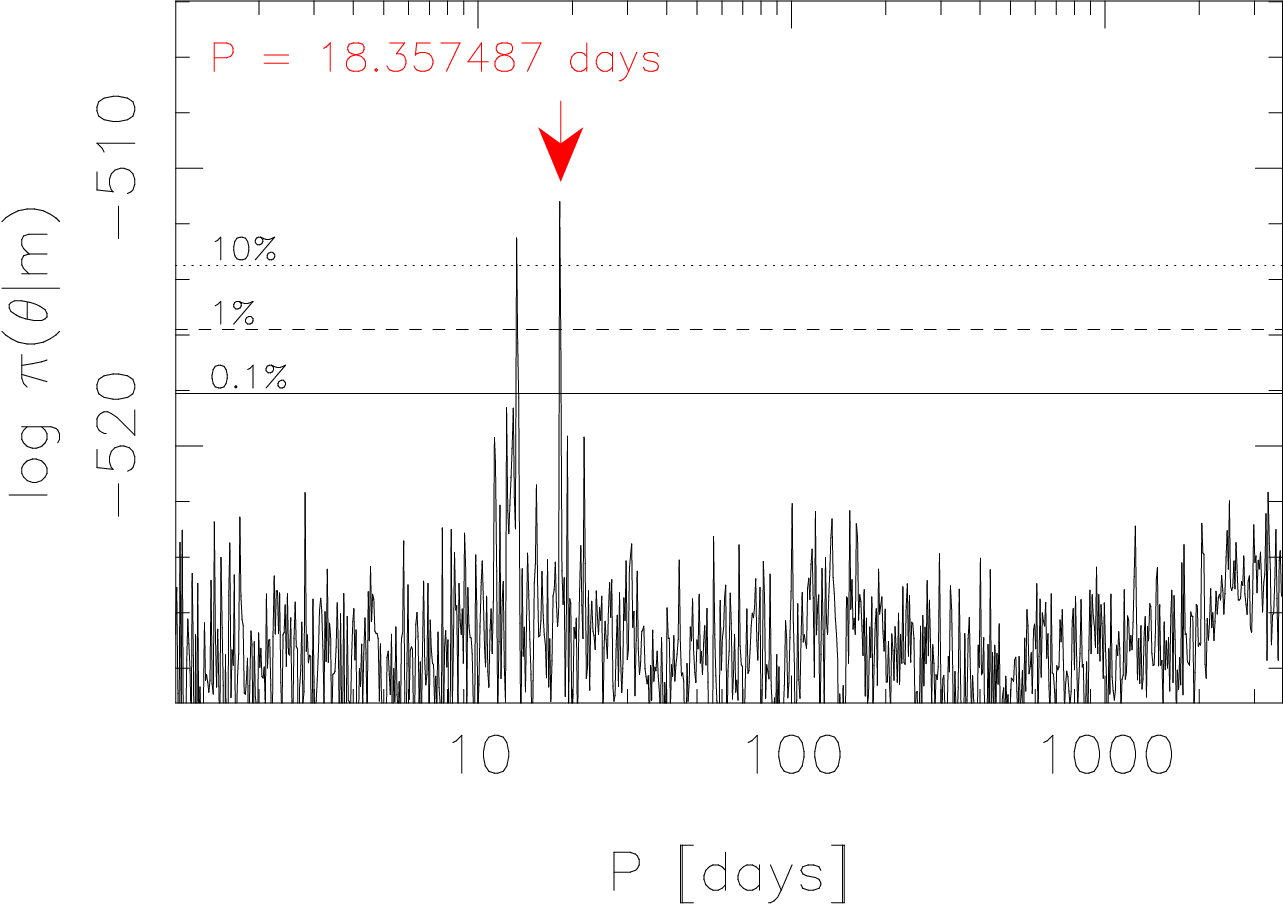}

\begin{minipage}{0.32\textwidth}
~
\end{minipage}
\includegraphics[angle=270, width=0.32\textwidth, clip]{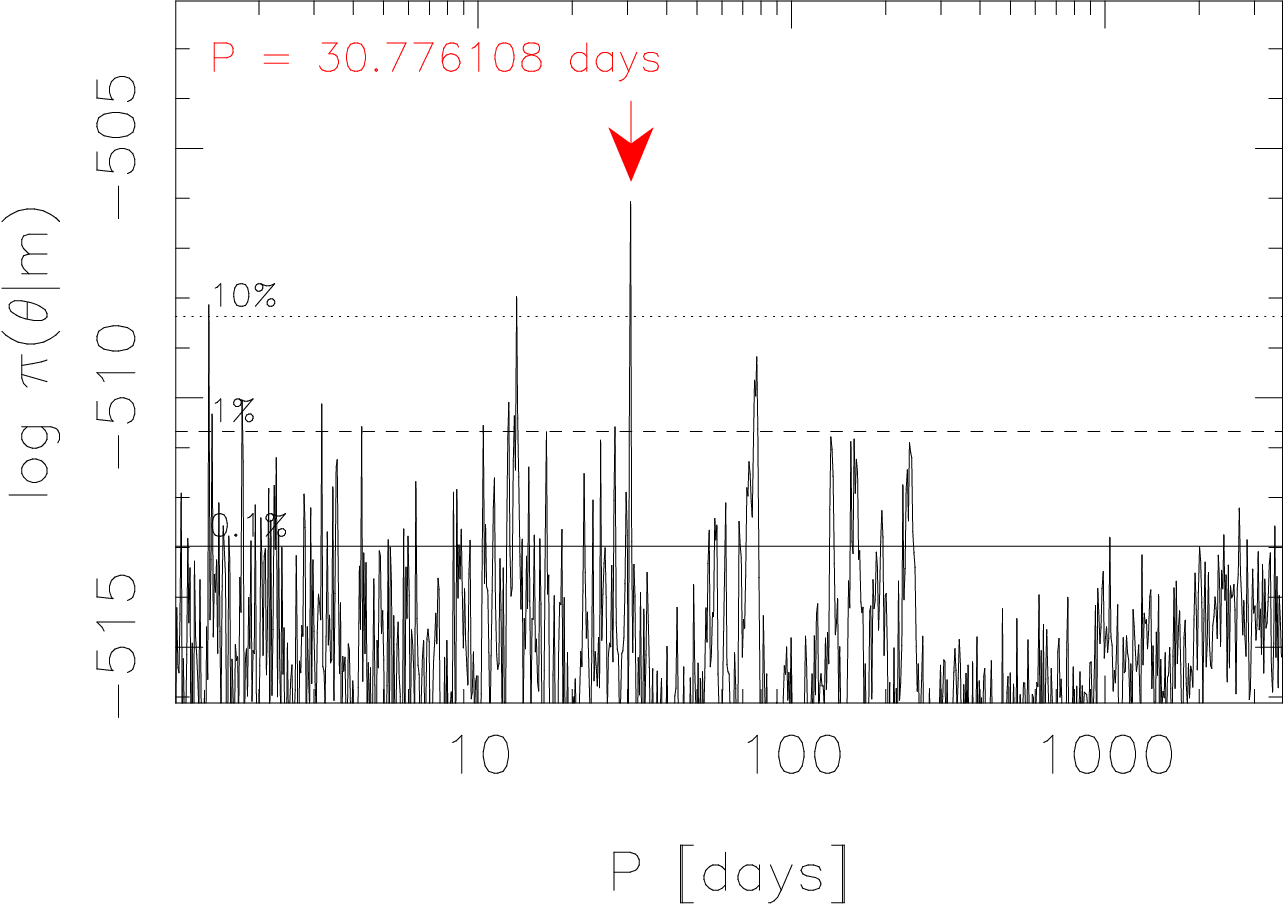}
\includegraphics[angle=270, width=0.32\textwidth, clip]{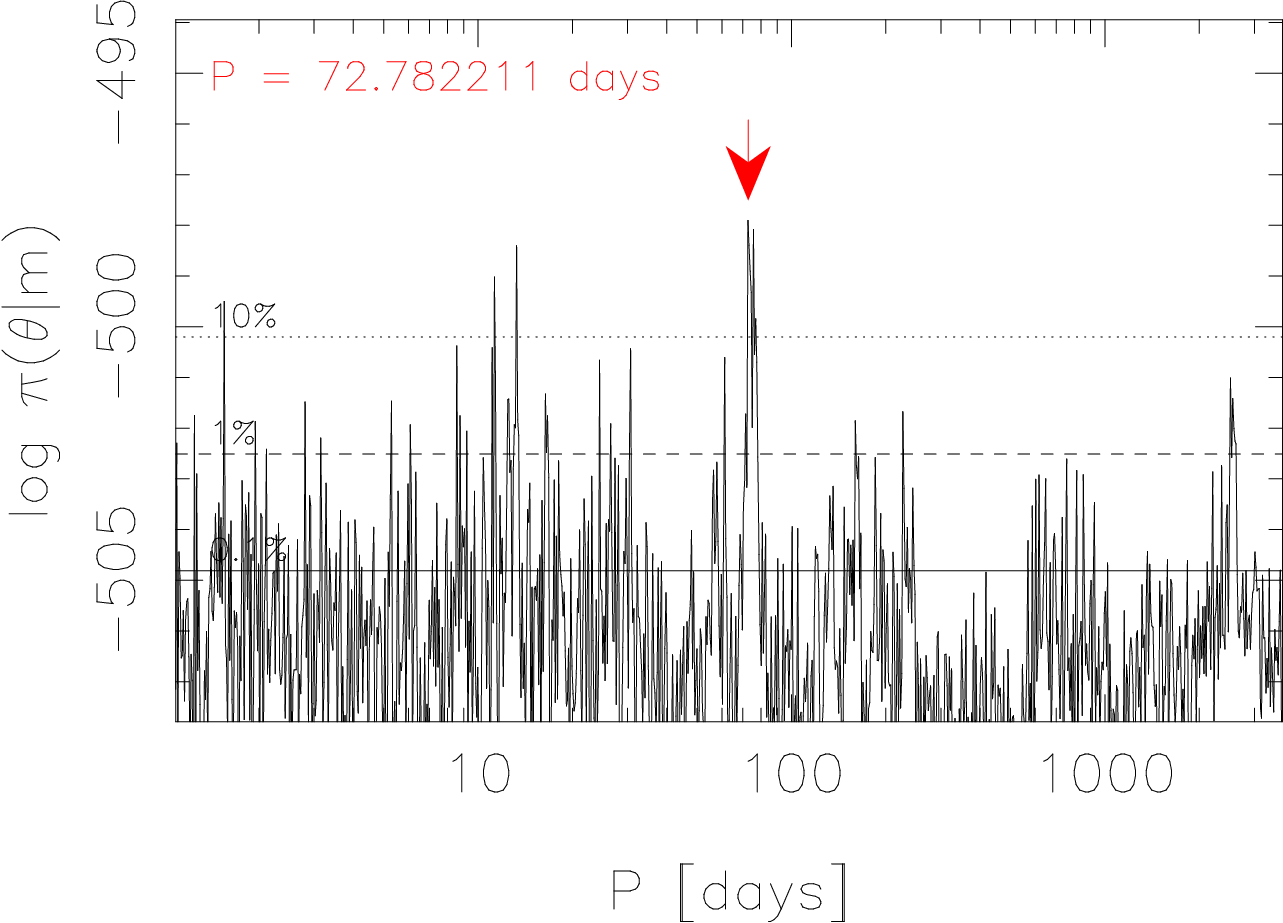}

\caption{Estimated posterior probability densities based on tempered Markov chain samplings as functions of signal period. From top to bottom we show the k=1,2, and 3 planet models. The maximum \emph{a posteriori} estimates identified by the chains are highlighted on the plots, as are the 0.1, 1, and 10\% equi-probability thresholds with respect to the maxima. The left column shows the pure white noise model, the middle column is for an MA1 red-noise model without activity correlation terms included, and the right column is for an MA(1) red-noise model that includes the activity correlations.  Note there is no k=3 planet model for the white noise model analyses (left column).}
\label{posteriors}
\end{figure*}

In the top panels of Fig.~\ref{posteriors} we show the posterior probability densities as functions of period for tempered Markov chain samplings employing three different $k=1$ signal (keplerian) models.  We can see that there appears to be three regions in the period space where the Markov chains identified considerable maxima for the pair of one-component moving average (MA(1)) models (middle and right columns) and only one region for the white noise model (left column). The most significant of these, i.e. the global maximum, was found to be at 25.6~days in all models.  The existence of the local maxima means we are likely to find other significant periodic signals in the data.

The panels in the middle row in Fig.~\ref{posteriors} show what we find when applying the $k=2$ signal models, the first one being at a period of 25.6~days. This time the most significant maximum is found to be at 18.35~days when dealing with the red-noise and also including the correlation terms, in excellent agreement with the pair of signals published previously in \citet{jenkins13b}. Therefore, we can confirm that there are two significant frequencies in the extended timeseries data for HD41248. The log-Bayesian evidences for these model comparisons can be found in Table~\ref{tab:evidence} listed as Full Data Set.  It can be seen that the white noise model search (left panel) did not find a unique second periodicity since it is masked by the correlated noise, variability related to activity, and increased jitter. \citet{santos14} could not confirm this second signal in their data and it is likely that this was due to inadequate noise modeling as they assumed a white noise model, fixed the excess white noise to an value of 0.7 \ms, and relied on analyses of model residuals that cause severe biases to the obtained solutions \citep{tuomi12,tuomi13c}.  We also note that the parameter density widths for both signals decrease significantly by including this new data, a feature not expected from a quasi-static source.  Furthermore, the linear trend applied to the data in \citet{santos14} is not significant, agreeing well with zero within any reasonable confidence level (Table~\ref{tab:evidence}). 

When employing the $k=2$ models a second significant peak around 13~days was found to cross the 10\% probability threshold in the analysis with activity correlation terms included, indicating a third signal could be present in the combined data.  \citet{santos14} also detected this signal and attributed it to the first harmonic of the 25~day signal, however our parameter densities suggest otherwise as the distribution did not overlap with one half of the period of the 25~day signal.  In any case, we then applied the $k=3$ models to test if this was indeed the case and we show the posterior diagrams in the bottom panels of Fig.~\ref{posteriors}.  No additional signals that were unique and passed our signal detection criteria were found in the analysis.  We did not employ the $k=3$ model to the white noise model as we could not constrain any secondary signal under that assumption.  

\begin{figure*}

\includegraphics[angle=270, width=0.5\textwidth, clip]{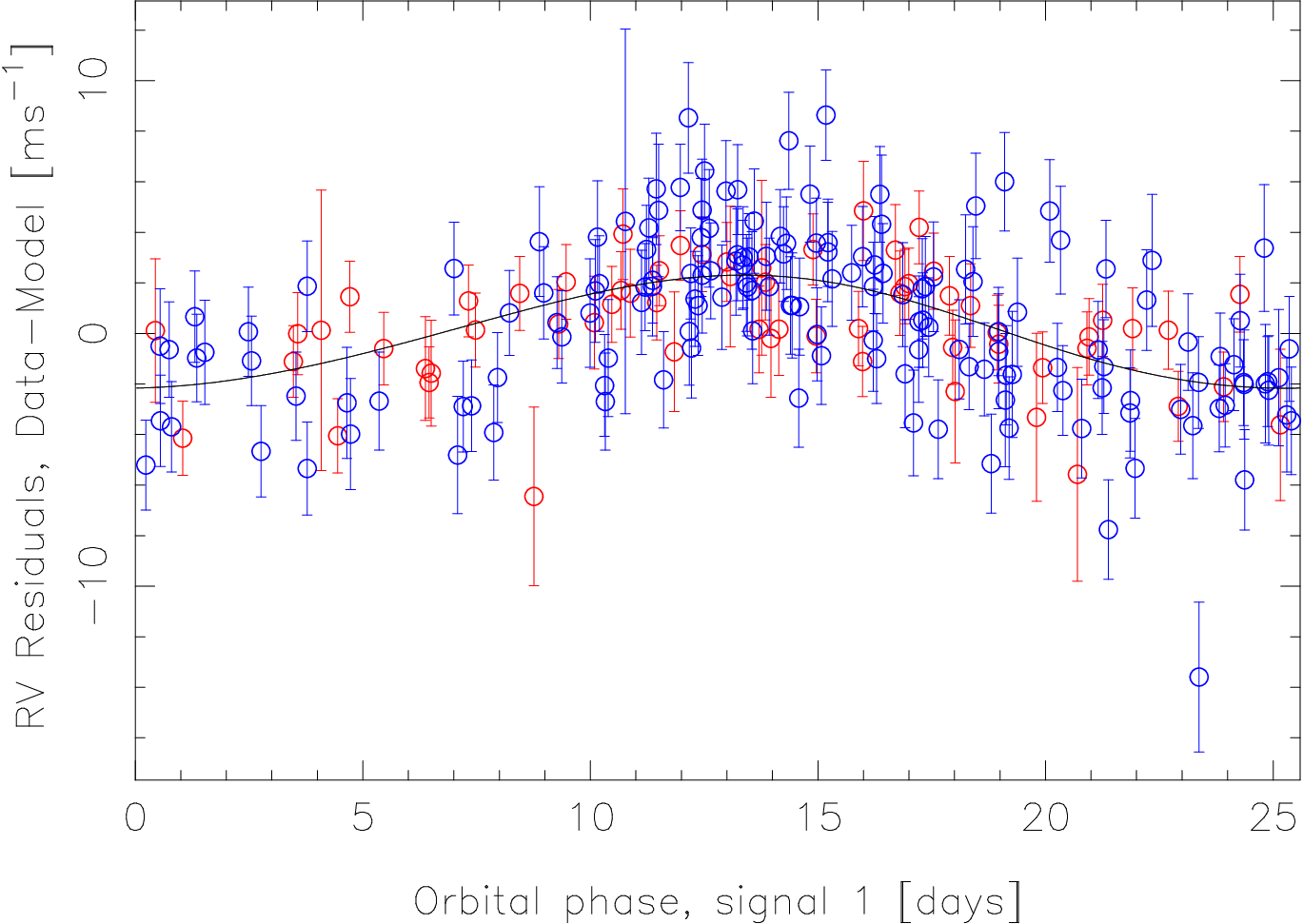}
\includegraphics[angle=270, width=0.5\textwidth, clip]{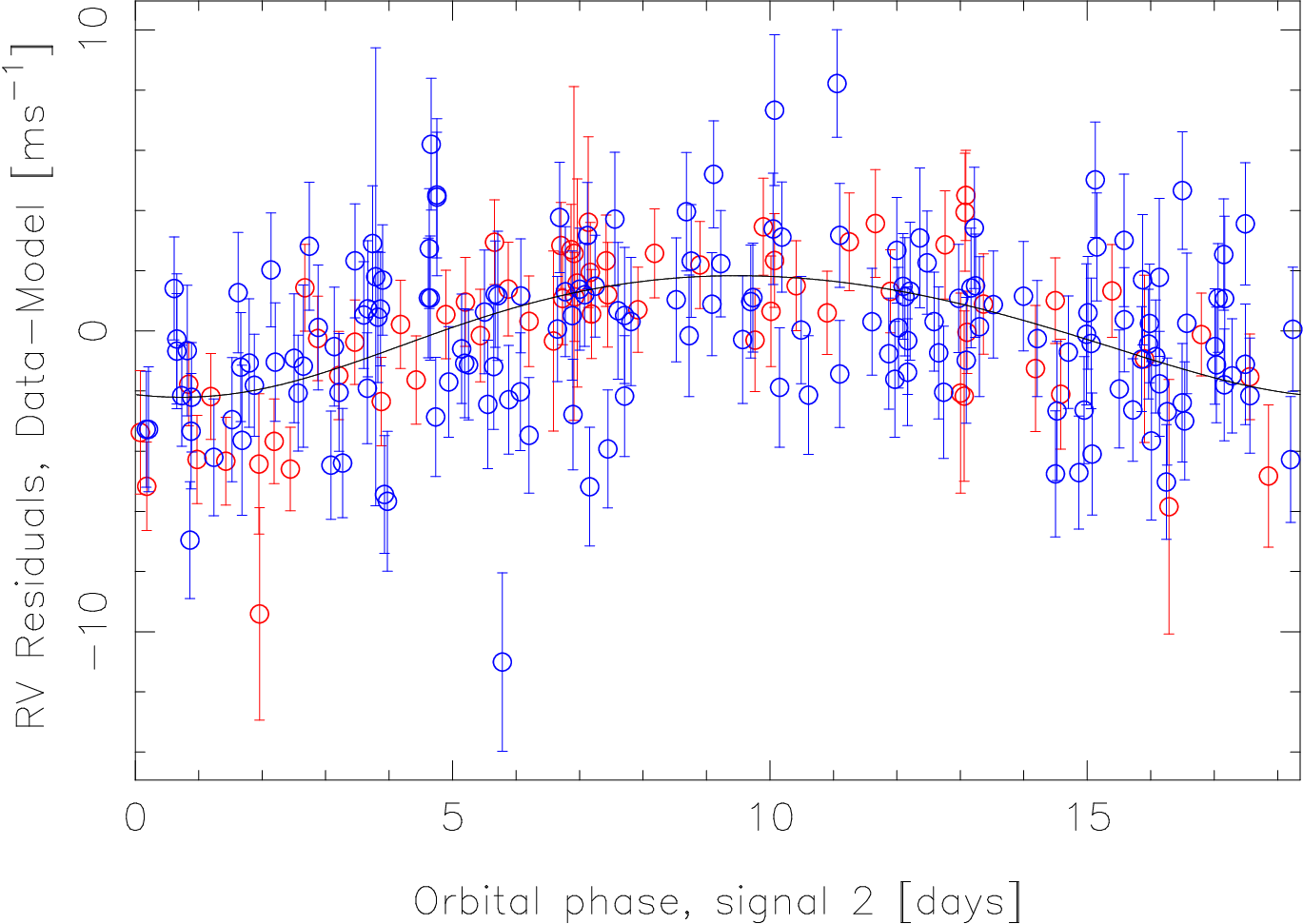}

\includegraphics[angle=270, width=0.5\textwidth, clip]{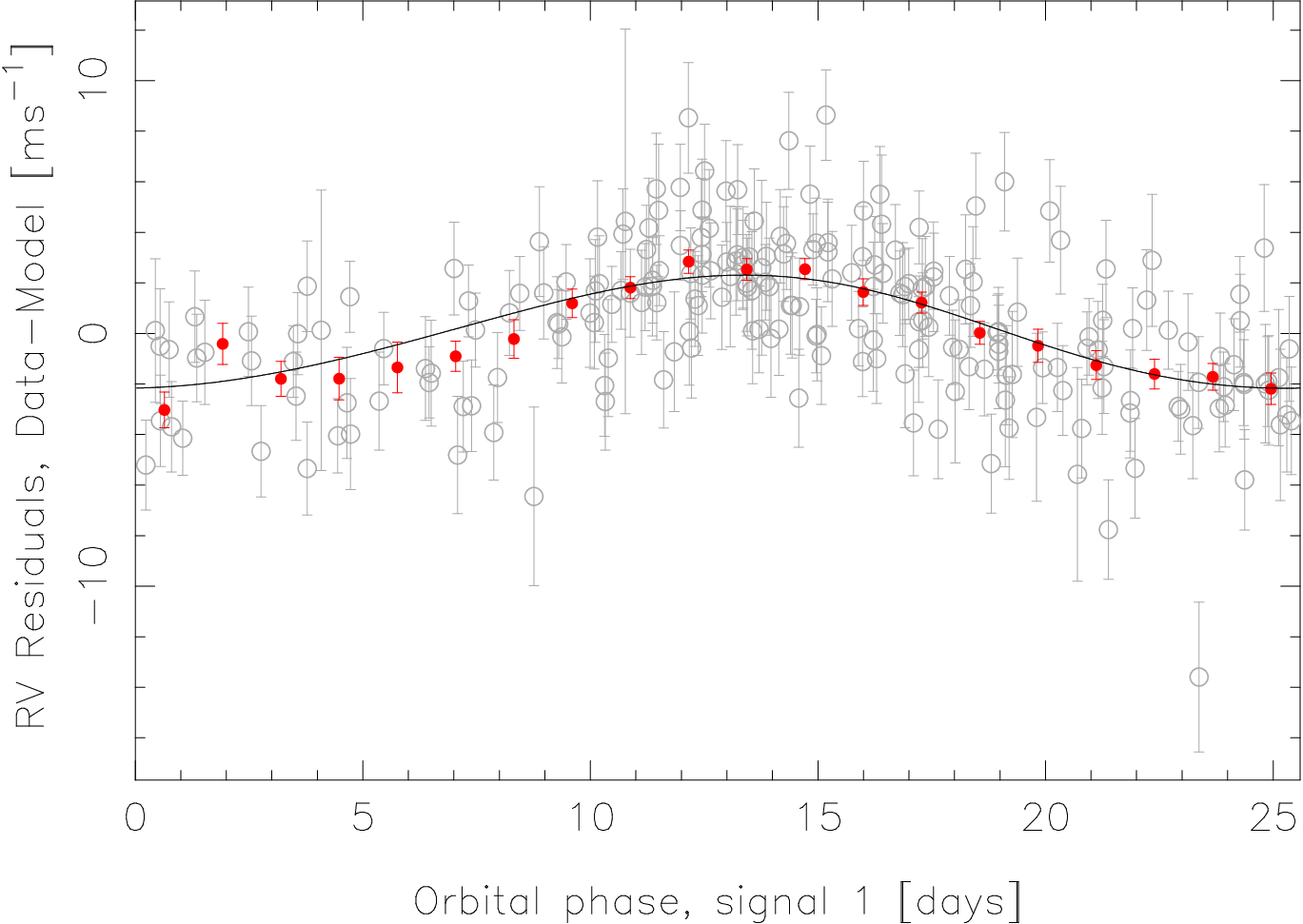}
\includegraphics[angle=270, width=0.5\textwidth, clip]{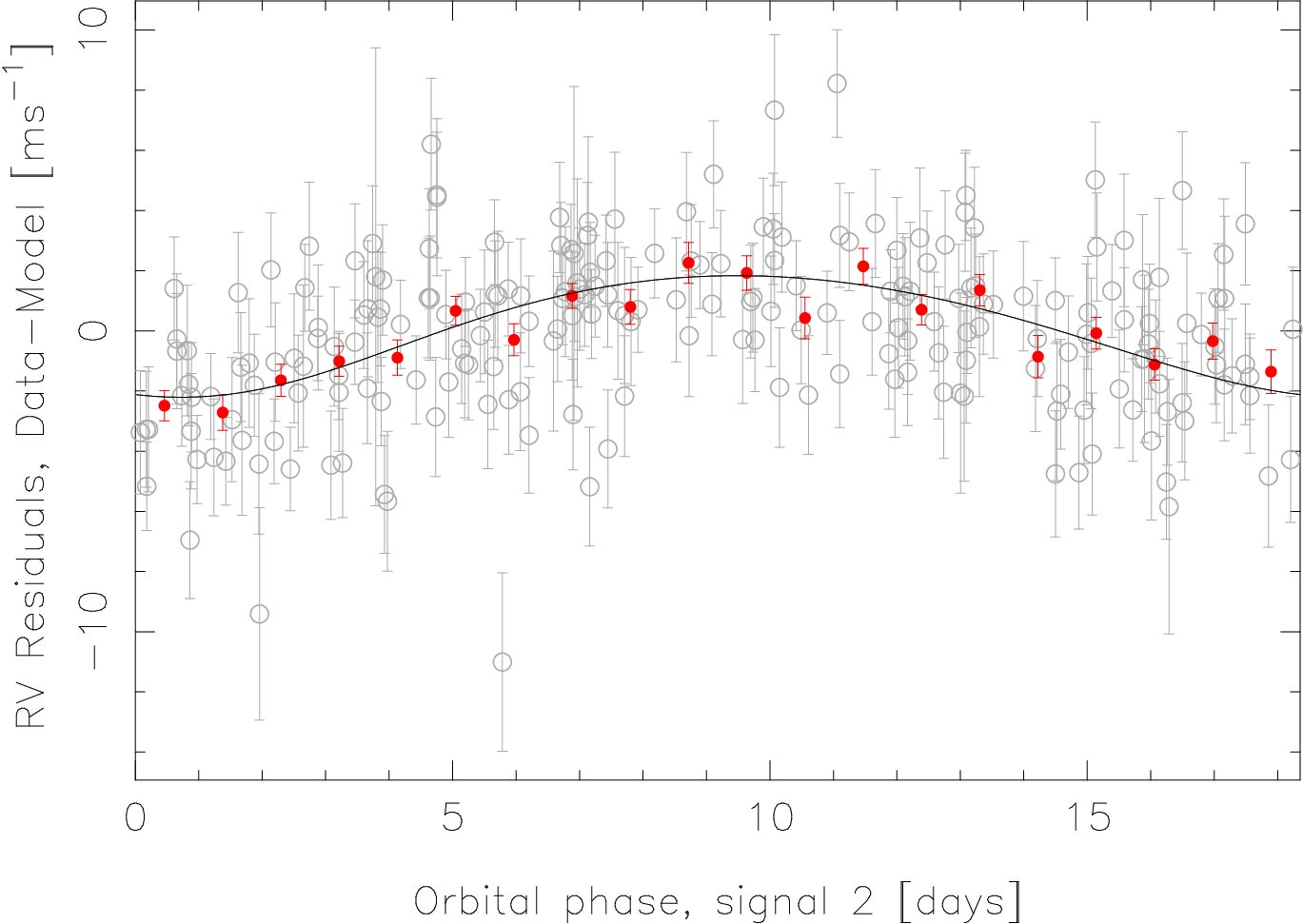}

\caption{Phase folded radial velocities for both signals detected in the HD41248 radial velocities as a function of time. The red points are for old low cadence data and the blue points are for new high cadence data in the top panels.  The lower panels show the same data but with filled red points representing binned velocities to highlight the significance of the signals.}
\label{time_analysis}
\end{figure*}

Given that no 13~day signal is found in the full data when including the noise correlations, and since the signal in the later data that we detect does not pass our planetary signal selection criteria, which are 1) the model including the signal must be 10000 times more probable than the previous model, i.e. $k+1 >> k$ and 2) the signal must not vary in time in period, phase, and amplitude over the baseline of the observations (see \citealp{tuomi14b}), it cannot be considered a static Doppler signal. This result is a reproduction of the same result found for the CoRoT-7 radial velocity timeseries (\citealp{tuomi14c}) where the rotation period of that star, known from the previously measured CoRoT photometry, did not correspond to a genuine velocity signal.  This strongly indicates that the 13~day signal reflects a quasi-static nature that changes as a function of time and could be the actual rotational period of HD41248 or a mix of the signal from the rotational period and differential rotation.  The final system parameters are listed in Table~\ref{tab:system} and log-Bayesian evidence ratios for these tests are shown in Table~\ref{tab:evidence}.  The phase folded signals are shown in Fig.~\ref{time_analysis}.

\subsection{ASAS Photometry}

A large part of the problem in the signal characterisation for HD41248 surrounds the star's rotational period. Therefore, in order to see if we could pin down the rotational period we searched the latest version of the ASAS photometric catalogue (\citealp{pojmanski97}) to see what useful data there is for HD41248 and if we could locate a plausible rotational period. We obtained $V$-band photometry for this star and after weeding out strong outliers (beyond 5-$\sigma$) and selecting only the best data, those classed as 'A' in the ASAS photometric grading, we were left with a total of between 420-650 photometric points processed through five ASAS $V$-band apertures.

Periodogram analyses of each of the aperture data revealed some significant peaks. A long period peak is found beyond 1000~days, likely attributed to the magnetic cycle of the star. Another cluster of peaks emerge around 200-300~days, peaks that also appear in the radial velocities under the assumption of pure white noise (see the left column in Fig.~\ref{posteriors}).  We found neither a statistically significant peak that matched the signals we see in the radial velocity data, nor a period that could plausibly relate to the stellar rotation period.

\subsection{Activity Periods}

In addition to analysing the photometry we also tested the activity indices by running both periodogram analyses and posterior samplings to constrain any frequencies in these indicators that would show activity cycles that could be the source of these radial velocity variations.  \citet{santos14} show possible correlations between the 25 day radial velocity period and similar periods in the \rhk and the CCF FWHM measurements.  

First we performed a periodogram analysis on the chromospheric activity indices, after removing 3$\sigma$ outliers from the sample that, when included only served to add noise.  This analysis revealed a statistically significant frequency with a period of 27.62~days, close to the radial velocity period for the primary signal in the data.  We then performed the same analysis on the FWHM values, using the same data set, and we found another statistically significant periodogram peak that matched the activity index peak, having a period of 27.93~days.  The period of this FWHM changed to 25.31~days when we included all the FWHM data and subtracted off a linear trend from the timeseries, closely matching the radial velocity signal.  The question to answer is whether these signals are related to the radial velocity variations.

In order to answer this question, we then ran MCMC samplings under our Bayesian approach to search for the signals independently and to constrain the significance of the signals and their possible extent in period space to see if they overlap with the radial velocity periods.  We found the global probability maximum in the cleaned \rhk indices to be located at 60~days with our samplings, closely followed by a 27.7~day maximum that matched the periodogram analysis.  However, neither of these signals could be detected in the data according to the signal detection criteria because their periods and amplitudes could not be constrained from above and below.  This means that we cannot rule out the possibility that these maxima are in fact statistical flukes caused by the combination of random noise and data sampling coupled with correlations.  

We then chose to perform the same analysis on the FWHM measurements and found the strongest frequency to be at $\sim$800~days, after considering the linear trend.  A cluster of probability maxima was found between 20-30~days in this analysis.  However, none of these maxima corresponded to a genuine signal because they did not satisfy our signal detection criteria.  Our interpretation was that the reason for the 25~day peak in the periodogram analysis was simply due to making the assumption that the noise is distributed in a Gaussian fashion, which does not appear to be the case as we find a significant red-noise component in both the $S$-indices and the FWHM measurements.  Therefore, the 25~day peak in the activity indicators is heavily model dependent.  This is in stark contrast to the radial velocity signal at 25~days which is found irrespective of the assumed noise model.

\subsection{Signal Coherence}

As discussed in the introduction, the signal found in the radial velocity timeseries of HD166435 only had a coherence time of around 30 days.  Combining this with the bisector correlations \citet{queloz01} ruled out the existence of this planetary candidate.  In order to test the probability that the proposed planet candidates HD41248$b$ and $c$ are real Doppler signals and independent from any activity correlations, we split the data up into two independent sets as a function of time.  The first set was the original data published in \citet{jenkins13b} but analysed using our current statistical model (dataset 1) and the second (dataset 2) is the new data that is around 2.5 times larger than the first data and was added in the analysis from \citet{santos14}. The full time baseline of data covers more than 10 years, where data set 1 spans over eight years and data set 2 covers two years, but at much higher cadence.

We analysed both sets independently and in combination and recovered two signals each time with high statistical significance.  The phase, period, and amplitude of the detected signals were in agreement with that published in \citet{jenkins13b}. Table~\ref{tab:evidence} shows the results for the sample for these analyses and includes results for the full data set with and without correlations with the activity indices.

The fact we see no change in the properties of the 25~day signal between the full data and the two subsets is remarkable because the star itself does change with time.  We found that the intrinsic noise, parameterised with the standard deviation of the excess white noise $\sigma_{\rm J}$, increased between the old and new datasets by roughly a factor of two.  For the old data, the jitter was found to be 1.3~\ms but for the new data the jitter increased to 2.6~\ms. Furthermore the data is more correlated in the second dataset, with the correlation parameter ($\phi_{\rm{1,}l}$) changing from being consistent with zero for old data to being significantly clustered around a value of 0.6 in the new data, meaning the noise becomes redder in the new data also.  This increase in stellar noise and correlation parameter means that the second signal at 18~days is no longer detected in the second set of data and it is likely the reason \citet{santos14} struggled to locate the 18~day signal as they did not account for these differences in the noise. This is also likely the reason they could not find a circular solution for the 25~day signal either.  We found a circular solution by both including the eccentricity prior or assuming a flat prior, however \citet{zakamska11} show that there is a bias towards higher eccentricities in radial velocity surveys, a bias that our eccentricity prior helps alleviate.  We also note that correlations between the activity indicators and the velocities became significant in the new data whereas these correlations were not significant in the old data set (see \citealp{jenkins13b}).

Although the 18~day signal cannot be independently detected in the new data, likely due to the fact that the star has become more active and therefore disabling the detection of this weak signal behind the increased jitter and red-noise, it is still well-supported by the new data because its significance increases considerably when comparing between the old data and the full data (see Table \ref{tab:evidence}).  The significance of the $k=2$ model increases by a factor of 2700 between the old data and the full data, where it is 1.8x$10^7$ times more probable than the $k=1$ model in the full data.  This means that this signal, together with the 25~day one, retain their properties throughout the data baseline and cannot be shown to be dependent on the changes in the stellar activity.  This result also shows that serendipitously observing the first epoch of data when the star was intrinsically more inactive and the noise was whiter allowed both signals to be confirmed in that data set.  There is the sampling cadence to consider here also, since in the first epoch of data the sampling density was lower than the later two observing seasons and it could also be that this decreased cadence helped to suppress the effects of the correlated noise. 

We then split the data further into three groups so that we could examine the data sets presented in \citet{santos14} directly.  We first consider the data published in \citet{jenkins13b} as one set as before (First Data), but then analyse the following two observing seasons individually, where the middle set in time runs from JD 55904.8-56414.5 and the later set runs from JD 56521.9-56632.7.  \citeauthor{santos14} claim that the 25~day signal amplitude evolves with time across these three observing epochs, a strong argument in favour of an activity induced signal, however we could not confirm this to be the case.  What we find is a complex evolution of the properties of the data due to changes in the star and sampling.  First off, we did not find any significant evidence for a change in the amplitude of the 25~day signal in the first and middle parts of the data, the amplitudes are in strong statistical agreement.  We could not confirm this result on the later data since we could not constrain any signal at all in this set and below we explain why.

As discussed previously, it appears that HD41248 became more intrinsically active throughout the timeseries of this data, in agreement with findings in \citet{santos14}, however not in a linear fashion.  The jitter noise increased in the middle part of the data significantly, going from 1.3~\ms in the first part of the data, up to 2.4~\ms in the middle part, and then dropped again to 2.1~\ms for the later data, although this drop is currently not statistically significant.  Added to this, the red-noise components change with time also, going from being consistent with zero in the first and middle parts of the data to a value of 0.6 in the later data, showing an increased red-noise in the later data as well.  

The activity correlations were also found to evolve throughout these three epochs of data.  Again these correlations were consistent with zero in the first part of the data, for all three indicators $S$-index, FWHM, and BIS, but apart from the $S$-index whose correlations with the velocities do not appear to evolve with time, the other two indicators do.  The correlations with FWHM ($c_{\rm{FWHM}}$) go from 0.13 in the middle data to 0.16 in the later data, both values statistically consistent but also statistically different from zero.  The correlations with the BIS go from being negatively correlated in the middle data, with a value of $c_{\rm{BIS}}$=-0.22$^{}_{}$, and then become in agreement with zero again in the later data.  Clearly the red-noise presents a complex pattern in this timeseries and when including all data after JD 55904.8, they are important and must be considered when searching for any low amplitude signals in this data.  In any case, it appears that our model does a reasonable job of describing the noise in the radial velocity timeseries for HD41248, similar to the case of CoRoT-7 (\citealp{tuomi14a}).

\citet{santos14} claim a stable period and phase could be maintained in HD41248 by an active longitude impacting the radial velocities (\citealp{berdyugina03}; \citealp{ivanov07}), under their hypothesis that the amplitude of the signal varies with time, which we have shown here cannot be concluded to be the case.  Yet \citeauthor{ivanov07} also show that although such an active longitude spot formation zones maintain their form much longer than the lifetimes of the individual spots, the solar data suggest they are only stable for 15-20 rotational periods.  We show that the 25~day radial velocity signal found for HD41248 has been static for \emph{at least} 5 years (70 rotational periods assuming the 25~day signal is the rotation period), and likely for the entire 10 year baseline of data (140 rotational periods).  

As a further test we decided to again split the data up into independent sets, this time based on their chromospheric activity levels, to see if we were sensitive to the impact of the increased magnetic activity and spot formation of HD41248.  We built three almost equally numbered sets, comprising an inactive set (\rhk $\le$ -4.93), an intermediately active set (-4.93 $<$ \rhk $\le$ -4.91), and an active set (\rhk $>$ -4.91).  We proceeded to search for the 25~day signal in these data sets and found that we could detect this signal in the inactive and intermediately active sets, but could not constrain anything in the active set.  This affirms why the 18~day signal cannot be detected in the later data and also calls for a noise model to be scaled as a function of chromospheric activity, an upgrade we plan to include in future versions of our model. This feature also highlights that we are sensitive to activity related features in our data and therefore, if the 25~day signal was genuinely due to rotationally modulated magnetic activity, we would expect the signal to appear stronger in the active data set than in the inactive data set, since the sensitivity to the features causing the signal increase.  We might also expect it to be more significant, depending on the structure in the increased jitter noise which would also be modulated by the rotation.  In any case, the signal parameters are invariant between the inactive and moderately active sets, showing that changes in the magnetic activity of the star do not change the period, amplitude, or phase of the signals, arguing against a magnetic origin for these signals.

Finally, we also tested the signals as a function of wavelength using the reddest HARPS orders only, (see \citet{anglada-escude12a} and \citet{tuomi13a} for details), and found no dependence of the signal properties or significances on wavelength. This indicates that neither of the signals show evidence for a dependence on wavelength, at least across the wavelength domain offered by HARPS. This would again argue against the origin of these signals being from magnetic activity cycles modulated by rotation.

\begin{table}
\center
\caption{Log-Bayes factors $\ln B_{1,0}$ and $\ln B_{2,1}$, i.e. in favour of one against zero and in favour of two against one-Keplerian models, given various divisions of the data and different models. The last column denotes the periods of the significantly detected signals. The models contain a moving average component. Two alternative two-Keplerian solutions are shown for the full data set.}
\label{tab:evidence}
\begin{tabular}{cccc}
\hline
\multicolumn{1}{c}{Data/Model}& \multicolumn{1}{c}{1} & \multicolumn{1}{c}{2}& \multicolumn{1}{c}{Periods} \\
 & & & [d] \\
\hline
Old Data & 18.4 & 8.8 & 25, 18 \\
New Data & 13.8 & -- & 25 \\
Full Data & 21.8 & 15.2 & 25, 18 \\
Full Data with Activity Correlations & 22.8 & 16.7 & 25, 18 \\
\hline\end{tabular}
\end{table}

\begin{table*}
\center
\caption{Solutions for HD41248.}
\label{tab:system}
\begin{tabular}{ccc}
\hline
\multicolumn{1}{c}{Parameter}& \multicolumn{1}{c}{HD41248~$b$} & \multicolumn{1}{c}{HD41248~$c$} \\ \hline
\hline
$P$ (d) & 25.595 [25.551. 25.652] & 18.361 [18.337, 18.392] \\
$K$ (ms$^{-1}$) & 2.30 [1.39, 3.21] & 1.95 [0.99, 2.83] \\
$e$ & 0.09 [0, 0.26] & 0.10 [0, 0.28] \\
$\omega$ (rad) & 0.3 [0, 2$\pi$] & 3.3 [0, 2$\pi$] \\
$M_{0}$ (rad) & 3.7 [0, 2$\pi$] & 5.6 [0, 2$\pi$] \\
\hline
$a$ (AU) & 0.166 [0.148, 0.180] & 0.132 [0.118, 0.146] \\
$m_{p} \sin i$ (M$_{\oplus}$) & 9.8 [5.9, 14.6] & 7.6 [3.6, 11.6] \\\hline
 & Old data & New data \\
\hline
$\tau$ (d) & 16.8 [0, 100] & 1.4 [0, 100] \\
$\phi$ & 0.17 [-0.47, 0.87] & 0.35 [0.10, 0.73] \\
$\sigma_{\rm J}$ (ms$^{-1}$) & 1.10 [0.77, 2.12] & 2.25 [1.74, 2.81] \\
c$_{\rm BIS}$ & -0.05 [-0.29, 0.19] & -0.11 [-0.28, 0.07] \\
c$_{\rm FWHM}$ & 0.03 [-0.09, 0.14] & 0.13 [0.02, 0.26] \\
c$_{\rm S}$ (ms$^{-1}$dex$^{-1}$) & 91 [-135, 343] & 67 [-93, 226] \\
\hline
$\dot{\gamma}$ (ms$^{-1}$year$^{-1}$) & 0.04 [-0.30, 0.37] \\

\hline
\end{tabular}
\end{table*}

\section{Summary}

We have shown that the radial velocity timeseries for the star HD41248, covering nearly 10~years of observation, clearly supports the existence of two signals with close to circular morphologies once red-noise components are considered.  This analysis provides additional evidence that the pair of signals detected in this data could be due to a pair of planets in, or very near to, the 7:5 MMR with periods of 18.361 and 25.595~days and a period ratio ($P_c/P_b$) of 1.394$\pm$0.005 at 99\% confidence level, in excellent agreement with the results published in \citet{jenkins13b}.  Such resonances are known to be a by-product of the planet formation and evolution process (\citealp{baruteau13}) in the early history of a star's life. It seems difficult to give rise to signals so close to such a period ratio simply by rotationally modulated activity in the presence of differential rotation, except in the most unique circumstances.

By analysing the signals as a function of time in an independent fashion we were able to obtain evidence for their static nature over the full baseline of observations. This analysis also allowed us to show that the star got intrinsically more active within the period of the most recent data, with the jitter noise taking on a value twice that reported in \citet{jenkins13b}. The noise also got significantly redder and the linear correlations with the velocities and FWHM increased such that they became statistically significant over the last two observing seasons.  Splitting the data further revealed a complex pattern of evolving red-noise and activity correlations, both of which would serve to mask weak signals under a white noise assumption.  We conclude that noise correlations must be taken into account when attempting to search for periodic signals that are at the noise level of the star/instrument combination.

We analysed the radial velocities with and without activity indicator correlations and found that both signals are supported by the old and new HARPS velocity data. Moreover, the significances of the signals increase when including the new data and when considering the activity indicator correlations. This is characteristic behaviour of a pair of static Doppler signals.  Including the linear correlation terms and red-noise correlations also results in removing spurious peaks from the white noise model search, peaks that appear in the activity indicators. We also show that the signals in the activity indicators are highly model dependent, only peaking at 25~days when gaussian noise is assumed, whereas the 25~day signal in the radial velocity measurements is found no matter what the assumption of the noise is.

Further tests revealed that we are sensitive to changes in the magnetic activity and we found that we could detect the 25~day signal in the radial velocities of the star when it was in its most inactive and moderatively active states.  No signals were detected when the star was in it's most active state, contrary to what would be expected if these signals were due to magnetic activity since we might expect these signals to be strongest when the star is in an active state as the source of the signals should give rise to a stronger signal. We also note that the signals are independent of wavelength in the band covered by HARPS, a further argument against the magnetic activity cycle argument.

Additional confirmation that these signals could represent a pair of resonant planets may have to wait until future instruments operating in the near infrared come online.  Instruments like HPF (\citealp{ramsey08}; \citealp{mahadevan12}) or CARMENES (\citealp{quirrenbach12}) could search for a change in the period or amplitudes of these signals as a function of wavelength over a much wider waveband than that offered by HARPS, which would attribute them to rotationally modulated spots on the surface of the star.  Future direct imaging systems like the previously proposed TPF or Darwin missions (see \citealp{leger00}), could be another way to confirm or not the existence of these planets, yet this type of mission is a long way off in the future and the distance of 52~pc to HD41248 makes this a real challenge.  In any case, more high cadence velocity observations over the coming years might be able to shed some light on the nature of these signals, either by searching for variations in the periods, amplitudes, and phases, or by confirming the nature of these signals with more high quality data.  HD41248 therefore represents a very interesting target to monitor radial velocity signals buried within evolving red-noise and activity correlations.

\acknowledgments

JSJ and MT acknowledge funding from CATA (PB06, Conicyt).  We also acknowledge the helpful and quick response from the anonymous referee.

\bibliographystyle{aa}
\bibliography{refs}

\end{document}